# Sub-wavelength surface IR imaging of soft-condensed matter

**James H. Rice,[1,3] Graeme A. Hill,[1] Stephen R. Meech,[1] Paulina Kuo,[2] Konstantin Vodopyanov,[2] Michael Reading[1]**

[1]*School of Chemical Science and Pharmacy, University of East Anglia, Norwich NR4 7TJ, UK.*

[2]*Ginzton Laboratory, Stanford University, Stanford, CA 94305-4088, USA.*

[3] *present address, School of Physics, University College Dublin, Belfield, Dublin, Ireland*

**Abstract**

Outlined here is a technique for sub-wavelength infrared surface imaging performed using a phase matched optical parametric oscillator laser and an atomic force microscope as the detection mechanism. The technique uses a novel surface excitation illumination approach to perform simultaneously chemical mapping and AFM topography imaging with an image resolution of 200 nm. This method was demonstrated by imaging polystyrene micro-structures.

**1. Introduction**

Infrared spectroscopy (IR spectroscopy) is a powerful and well developed method to characterise the chemical composition of materials. IR spectroscopy utilises infrared radiation commonly from a black body emitting source to tune across vibrational absorption frequencies [1]. IR is frequently applied to characterise molecular species by utilising the characteristic vibrational modes (eigenmodes) that each molecular species possess.

Microscopy based methods have made great progress in developing techniques that can image on the nanoscale, these approaches can be divided into two main classes, near-field and far-field techniques. IR Imaging using far-field methods is limited by diffraction to length scales of the order of 1-5 micrometers for IR microscopy [2]. It is noted that fluorescence based imaging has successfully over come the diffraction limit in the far-field to enable an image resolution as good as tens of nanometres to be archived [3], however to date no progress has been reported for far-field IR based imaging.

Scanning probe techniques such as Atomic force Microscopy (AFM) enable topographic imaging of surfaces with very high spatial resolution, i.e. < 10 nm [4]. Such an approach to imaging is a well developed near-field method. Combining IR spectroscopy with AFM potentially allows for a method that can chemically map materials with nanometre spatial resolution. Such an approach based on combining AFM and IR spectroscopy has been developed [5-18].

IR spectra of analytes and thin films have been recorded by combining a Fourier transform infra-red spectroscopy instrument with an AFM [5,6]. This approach is based on an opto-thermal method that utilizes commercial AFM probes as temperature sensors which enabled measurements of opto-thermal signals induced by absorption of IR radiation [7,8]. This method enabled measurements of samples down to 5 μms in size. An alternative approach to opto-thermal based imaging, referred to as AFMIR, has enabled IR imaging of materials with sub-wavelength lateral resolution [9-12]. This method measures IR absorption directly via measuring local transient deformation in the AFM cantilever induced by an infrared pulsed laser tuned at a vibrational absorbing wavelength generating IR absorption spectral information.

AFMIR has been applied to image E. Coli bacteria at N-H resonance frequencies, expitaxial quantum dots in resonance with intra-sub band transitions and live cells monitoring the glycogen band centred at 1080 cm$^{-1}$ with a sub-diffraction limited image resolution of 60-100 nm [10-12]. This experimental methodology requires the use of an attenuated total internal reflection (ATR) arrangement in combination with IR cyclotron radiation. This has effectively limited the application of the technique to samples of thicknesses on the order of the excitation wavelength due to the requirement of evanescent wave propagation through the sample. In order to study samples which are thicker alternative experimental designs are required. Here, we outline an experimental method for sub-micron IR surface imaging (referred to as s-AFMIR) that directly excites samples surface via a top-down excitation arrangement [13]. Additional novel aspects of the outlined experimental approach include the use of an optical parameter oscillator (OPO) tuneable IR source, replacing cyclotron and black body radiation filaments as sources for IR radiation.

This approach is different from other work on work combining AFM and optics to enable sub-diffraction IR imaging. Hillenbrand et al applied a method referred to as scattering-type scanning near-field optical microscopy (s-SNOM) to enable sub-diffraction IR imaging [14-16]. In s-SNOM the metalized tip of an Atomic Force Microscope (AFM) is illuminated by laser light. The effect of this is to excite localized surface phonon-polaritons at the tip apex by the concentrated optical field. This produces a near-field resonance at a material specific frequency close to the LO phonon frequency. This method requires the AFM tip to oscillate and the scattered light is collected in the far-field. Aigouy et al also has applied a oscillating tip and far field light collection methodology to enable sub-diffraction optical imaging [17-18]. The outlined method here contrasts with these approaches to perform sub-diffraction imaging by using the excitation light field to induce a topographic change in the sample, the intensity of which is measured as a function of wavelength in in order to construct an IR spectrum. This requires that the AFM tip be keep in contact with the sample and there is no need to collect the scattered light from the sample. This enables the measurement directly of light absorption. The scattering experiment provides less direct access to the IR absorption spectra of materials as the measured response is a convolution of this and the local dielectric properties.

## 2. Instrumentation

The s-AFMIR experimental set-up was based on an in-house built IR laser source and a commercial Atomic force microscope. Mid-IR radiation was generated using a periodically poled $LiNbO_3$ crystal emitting tuneable IR laser radiation, tuneable over 3.13 to 3.57 μm as demonstrated by Vodopyanov [19]. The OPO laser provides a high power infrared light source (compared to black-body sources). The OPO was constructed using an PPLN crystal (Super Optronics Inc) and a nanosecond Nd:YAG pump laser (model NL202, EKSPLA). The PPLN crystal possessing a period of 29.0-30.6 μm fanning and was kept at a constant temperature of 90 $^0$C using a crystal oven heater (temperature controller, model TC038 and model III crystal heater, HC Photonics Corp). The OPO set up is outlined in Fig 1. Selective filtering of laser radiation was performed using a long pass filter (L.P-2500 nm, Spectrogon). Wavelength output was monitored using the frequency mixed component generated at the 800 nm region of the visible spectrum using an ocean optics fibre coupled CCD and spectrograph. The PPLN crystal was mounted onto a linear stage (Melles Griot) and was scanned with the wavelength output monitored. The output power was monitored to be c.a. 3 mW with pump power of c.a. 400 mW.

The IR laser is directed onto the surface of the sample been probed by the AFM tip using mirrors as outlined in Fig 1. The excitation beam was focused using a convex mirror. Careful positioning of the excitation beam with respect to the tip was required to optimise the sample. The AFM (Veeco Explorer system) was used with a scanner with lateral and vertical scans of 100 x 100 μm and 10 μm respectively. A Stanford SR650 programmable filter was used to amplify and filter the signal from the AFM. The filtered signal was routed to the input of an Agilent DSO5012A oscilloscope where it was retrieved using in-house software and analyzed with the WSxM program [20]. Tips were silicon nitride v-shaped cantilever tips (used as received from Veeco). The AFM is operated in contact mode, with a typical setpoint of the tip between 5 and 10 nN. Through analysis of the force curve of the contact and by establishing the setpoint, the AFM can remain in a linear range in contact with the surface in line with the literature [10].

Samples were prepared on CaF2 (Aldrich) or Mica substrates. Analytical grade glycerol (Aldrich), polystyrene (PS) films (Aldrich) and 3 μm sized beads (Aldrich) were used. Polystyrene beads (PSB) were dispersed onto a mica surface and heated. This enabled the beads to be fixed to the substrate and are not free to be moved when interacting with the probing AFM tip. PS film was mounted directly onto the substrate. The PSB (on average 3 μm in diameter) were purchased from Aldrich. A commercial Attenuated Internal reflection infrared (ATR-IR) spectrometer was used to record a reference IR spectrum of polystyrene.

## 3. Origin of s-AFMIR signal

The origin of the s-AFMIR comes from optically induced effects that are directly proportional to IR absorption. The AFM cantilever tip was positioned over the sample with the tip in contact with the sample surface (as outlined schematically in Fig 2). Following the application of the IR radiation source the cantilever tip response was monitored. The specific response studied was a change in the vibrational motion of the tip arising from absorption of radiation by the sample. The energy absorbed is dissipated through thermal and acoustic mechanisms. Propagating acoustic waves create a deformation in the surface topography which can be detected by the AFM tip (see Fig 2). As an IR laser excitation source is tuned into resonance with a vibration mode, absorption of IR radiation increases and in turn the energy absorbed increases. Recording the cantilever displacement intensity as a function of excitation wavelength enables either IR spectra or IR imaging to be performed when the AFM cantilever is stationary or is scanning respectively. The absorption of the excitation radiation exponentially falls as the excitation radiation propagates through the absorbing material [21,22]. As a consequence the applied energy from the excitation radiation will be converted into thermal and acoustic energy predominantly at and near the surface of the sample. The effective spatial resolution been a function of the sample under study. AFMIR has been applied to study quantum dots with a spatial resolution of 60 nm [11]. This indicates that the propagation lengths of the acoustic waves that are responsible for the absorption measurement mechanism are short enough in such materials to enable nanoscale imaging.

## 4. Studies of bulk material

Polystyrene films (20mm x 20 mm x 1 mm) were studied using s-AFMIR. The AFM cantilever tip was positioned over the centre of the PS film with the tip in contact with the sample. Following the application of the IR radiation source the cantilever tip response was monitored. Fig 3 shows an s-AFMIR spectrum recorded for a PS film, shown along with an IR spectrum recorded using a commercial IR instrument. Inspection of the spectra show that both possess the same spectral features. This indicates that the s-AFMIR method has recovered the IR spectrum for a PS film with the characteristic vibration intensities arising from the aromatic ring, $(C_6H_5)$, giving rise to the group of bands above 3010 cm$^{-1}$ and bands below 3010 cm$^{-1}$ present arising from the saturated main chain CH groups.

## 5. Mesoscopic systems

### 5.1. Localised spectroscopy

Studies of individual PSBs were undertaken to demonstrate imaging small microscopic objects using this approach. A single PSB, 3.5 μm in diameter and 750 nm in height was located using AFM topography scanning. The AFM cantilever tip was positioned over the centre of the PSB with the tip in contact with the sample. Recording an s-AFMIR signal when the AFM tip is stationary enables an IR spectrum of a sample to be recorded as shown in Fig 3. The spectrum is shown along with a reference IR spectrum of a polystyrene film, recorded using a commercial IR instrument for comparison.

The s-AFMIR spectra shown in Fig 3 possess a wavelength resolution of 12 cm$^{-1}$, while the ATR-IR based spectrum has a wavelength resolution of 1 cm$^{-1}$. The s-AFMIR resolution was due to the scanning increments. It is noted that the limiting resolution will be the PPLN IR source which is on the order of <6 cm$^{-1}$ with the outlined experimental set up. In order to access this resolution the use of a precision controlled motorised stage or manual stage with micron resolution is required to scan across the PPLN crystal across the Nd:YAG pump laser beam. This must be coupled with precisely focused laser pump excitation beam that matches the resolution in the fanning gradient in the PPLN crystal, noting that the pump power must not exceed the damage threshold of the PPLN crystal.

Analysis of the s-AFMIR spectrum of PSB in Fig 3 shows that there are differences in the spectral features when compared to the IR spectrum of the PS film. The band at 2960 cm$^{-1}$ is larger in relative intensity in the PSB spectrum compared to the PS film.

This spectral feature difference may arise from the local structural form that each of the polymers possess. Studies of PS with different morphologies has shown that the 2960 cm$^{-1}$ band is sensitive to the presence of these morphology differences i.e. spherical beads or linear films of PS, it is noted that the heating of the PSB beads may also result in the formation of different polymer phases resulting in the observed relative intensity changes [23,24].

The AFMIR spectrum was recorded using the FFT of the cantilever oscillation as a function of excitation wavelength. The FFT of the signal shows the presence of numerous peaks corresponding to the different resonant vibrational modes of the AFM tip as a function of frequency. By monitoring the intensity of the FFT peaks it is possible to track the resonant response of the cantilever that is influenced by the absorption behaviour of the sample. A previous report by Hill et al outlined that s-AFMIR imaging showed that through the application of FFT signal processing the signal to noise ratio can be increased [13]. The use of FFT signal processing creates a negligible background for the off-resonance signal making top-down s-AFMIR and ATR AFMIR methods comparable in regard to reduction in off-resonance background signal which demonstrates the high contrast of spectroscopic measurement capabilities in both approaches.

The image resolution of s-AFMIR was estimated to be 200 nm. Measuring a line scan across the edge of a PSB and comparing the s-AFMIR resolution with an AFM topography scan recorded at the same time estimated the s-AFMIR resolution to be 200 nm comparable with that of the AFM (see Fig 4). As Fig 4 and Fig 4's insert at a spatial separation of 200 nm the AFMIR signal be differentiated from each proceeding point producing a line-scan plot that was almost identical to that of the AFM.

## 5.2. Sub-wavelength imaging

Fig 5 shows IR images of a single PBS. Shown is an AFM topography image and IR images at two different excitation wavelengths. The IR images were recorded by monitoring the FFT signal as a function of excitation wavelength when the AFM tip was been scanned. AFM topography images were recorded simultaneously. The IR image was constructed by recording an s-AFMIR spectrum per image point thus building an IR adsorption image. Fig 5 shows an s-AFMIR image recorded of the PSB at an on-resonance excitation wavelength. The white pattern is assigned to arise from IR absorption originating from a strong opto-acoustic oscillation response by the PSB, an absorption which is absent for the mica substrate. Comparing the AFM topography and s-AFMIR images shows good agreement in the image features. Fig 5 also shows an s-AFMIR image recorded of the PSB at an off-resonance excitation wavelength. The s-AFMIR shows a reduced white background in line with reduced IR absorption. 3D plots of s-AFMIR images of a single polystyrene bead are shown in Fig 5. 3D plots enable the analysis of the different IR absorption intensity for the s-AFMIR images on and off resonance to be clearer. 3D plot of an on-resonance s-AFMIR image at 2920 cm$^{-1}$ is shown along with a 3D plot of off-resonance excitation recorded at 2065 cm$^{-1}$. Inspection of the two 3D plots shows clear differences in signal intensity. The Plots recreate the physical dimensions of the polystyrene bead as measured with AFM topography methodology with the greater polystyrene s-AFMIR signal coming from the centre of the bead, where AFM topography measurements show the greatest polystyrene mass to be present. Differences between AFM topography and s-AFMIR images which are present may arise from the presence of thermal drift of the mechanical device occurring during the duration of the measurement of s-AFMIR images i.e. 20 min. Analysis of the IR images shows that the left hand side of the PSB is more pronounced than the right hand side. This is potentially associated with the angle of incidence of the excitation laser. Previous AFMIR studies of E.coli [10] reported that the resultant IR images of the bacterium possessed weak intensity at centre of the bacterium and relatively intense signal at the edges. This was explained due to the bacterium scattering the incident electric field around the bacterium resulting in a lightening distribution and not a homogenous distribution. In additional it was noted that the signal is greater at one side than the other which was explained by referring to the angle of incidence of the excitation light. In order to overcome these inhomogeneous distributions requires careful choice of the angle of incidence of the excitation radiation beam. Never the less in the reported case, and in the images presented here, the IR images match very well with the AFM topography images. Analyses of smaller sample sizes were undertaken. Fig 5 shows an AFM topography image and an s-AFMIR image recorded on-resonance at 2965 $cm^{-1}$ of moulded PSB formed by heating the sample. The AFM topography image shows a maximum height of 367 nm for the sample. The s-AFMIR image replicated the AFM topography features and demonstrates that thin surface features on the order of >367 nm can be imaged.

Fig 6 shows the corresponding IR absorption images of the same bead recoded at five different wavelengths. These images were recorded at a 5x times faster image rate than images shown in Fig 5. The IR absorption image recorded on-resonance show an intense white pattern present on a dark background. The white pattern is assigned to arise from IR absorption originating from a strong opto-acoustic oscillation response by the PSB, an absorption which is absent for the mica substrate. The IR absorption image recorded after moving the excitation wavelength off resonance shows the presence of a much weaker white pattern. This is expected as the IR absorption is much weaker but still present, resulting is a weaker but still visible white pattern. Moving further of resonance results in an image with no white pattern visible. While the rapid accumulation of the images results in distortion of the s-AFMIR image compared to those in Fig 5, it allows for relatively rapid analysis of the presence of polystyrene. Inspection of Fig 6 shows a correlation of peak intensity in the localised s-AFMIR spectrum with intensity of the presence of the PSB, as denoted by the intensity of the light contrast colouring. The strongest contrast is seen at the peak of the vibrational band. Imaging on the side band shows an image with a white pattern of intermediate intensity in line with the intensity of the side band seen in the s-AFMIR spectrum shown in Fig 6.

## 6. Potential improvements

Top-down excitation arrangement enables adequate excitation of a surface area of several microns using the out-lined set-up. The focusing of the excitation IR wavelengths via the use of an elliptical mirror practically enables focusing of the IR beam on the order of >10 microns. Focusing at the diffraction limit of $\lambda/2$ would enable a beam on the order of two microns to be achieved. This would limit s-AFMIR imaging to length scales on this order as imaging must occur within the IR excitation area. Imaging was performed without tightly focusing the excitation beam in order to excite a large area. This was done at the expense of signal sensitivity which would expect to improve with higher laser excitation power.

A number of improvements in the experimental system may lead to a further improvement in image resolution. Studies have shown that the tip motion is sensitive to both tip-surface and the tip-environment interactions [9]. This indicates that choice of tip composition should consider the material system been probed in order to optimise these interactions. This may include optimisation of AFM tip design to lead to more sensitive cantilever response in addition to reducing signal response arising from direct AFM tip excitation by the excitation laser. Studies of the effect of temperature on microcantilever resonance response showed that for mono-material cantilevers frequency varies in direct proportion to the temperature in line with the decrease in Young's modulus with increasing the temperature. When the cantilever is bi-material, the response is nonlinear due to differential thermal expansion. This indicates the choice of material used to build the tip with is important. Optimisation of AFM tip design may also include introduction of opto-thermal functionalism taking advantage of potential research advances in the development of opto-thermal IR imaging which will increase the range of measurements that can be made using the experimental approach outlined here.

**Conclusion**

A technique for sub-wavelength infrared surface imaging and spectroscopy of micro-structured polymeric material has been demonstrated with an image resolution of $\lambda/100$. In this way the surface features of a sample can be chemical characterised via inspection of the resulting IR spectra and images. This was achieved by combining a phase matched optical parametric oscillator laser as the IR source and an atomic force microscope as the detection mechanism. The technique uses a novel surface excitation illumination by focusing, at the diffraction limit, the excitation light directly onto the surface of the sample in presence of the AFM tip. This method enables chemical mapping and AFM topography imaging to be performed simultaneously with an image resolution of 200 nm.

**References**

1.  B. H. Stuart, *Infrared spectroscopy: fundamentals and applications* (John Wiley 2004).

2.  E. Betzig, J. K. Trautman, T. D. Harris, J. S. Weiner and R. L. Kostelak, Science, **251**, 1468 (1991).

3.  J. H. Rice, Mol. BioSyst., **3**, 781 (2007).

4.  T. R. Albrecht and C. F. Quate, J. Vac. Sci. Tech. A., **6**, 271 (1988).

5.  A. Hammiche, M. H. Pollock, M. Reading, M. Claybourn, P. M. Turner and K. Jewkes, Appl. Spectrosc., **53**, 810 (1999).

6.  A. Hammiche, L. Bozec, M. J. German, J. M. Chalmers, N. J. Everall, G. Poulter, M. Reading, D. B. Grandy, F. L. Martin and H. M. Pollock, Spectrosc., **19**, 20 (2004).

7.  A. Hammiche, L. Bozec, H. M. Pollock, M. German and M. Reading, J. Microsc. (Oxf), **213**, 129 (2004).

8.  M. Reading, D. M. Price, D. B. Grandy, R. M. Smith, L. Bozec, M. Conroy, A. Hammiche and H. M. Pollock, Macromol. Symp., **167**, 45 (2001).

9.  A. Dazzi, R. Prazeres, F. Glotin and J. M. Ortega, Infr. Phys. Tech., **49**, 113 (2006).

10. A. Dazzi, R. Prazeres, F. Glotin and J. M. Ortega, Ultramicroscopy, **107**, 1194 (2007).

11. J. Houel, S. Sauvage, P. Boucaud, A. Dazzi, R. Prazeres, F. Glotin, J.M. Ortega, A. Miard, A. Lemaıtre, Phys. Rev. Lett., **99**, 217404 (2007).

12. C. Mayet, A. Dazzi, R. Prazeres, F. Allot, F. Glotin, and J. M. Ortega, Optics Lett, **33**, 1611 (2008).

13. G. Hill, J.H. Rice, S.R. Meech, P. Kuo, K. Vodopyanov, M. Reading, Optics Lett., **34**, 431 (2009).

14. J. Aizpurua, T. Taubner, F. J. Garcia de Abajo, M. Brehm , R Hillenbrand, Opt. Expr. **16**, 1529 (2008).

15 . Cvitkovic, N. Ocelic,  R. Hillenbrand, Nano Lett., **7**, 3177 (2007).

16. M. Brehm, T. Taubner, R. Hillenbrand, F. Keilmann, Nano Lett. **6**, 1307 (2006).

17. F. Formanek, Y. De Wilde,  L. Aigouy, Ultramicroscopy **103**, 133 (2005).

18. A. Fragola, L. Aigouy, P.Y. Mignotte, F. Formanek, Y. De Wilde, Ultramicroscopy **101**, 47 (2004).

19. K. L. Vodopyanov and P. G. Schunemann, Opt. Lett., **28**, 441 (2003).

20. I. Horcas, R. Fernandez, J. M. Gomez-Rodriguez, J. Colchero, J. Gomez-Herrero, A. M. Baro, Rev. Sci. Instrum., **78**, 013705 (2007).

21. H. T. Jung, Appl. Phys. B **70**, 237 (2000).

22. H. N. Subrahmanyam, S.V Subramanyam, J. Mater., Sci., **22**, 2079 (1987) .

23. J. Duchet, S. Demoustier-Champagne, Polymer, **41**, 1 (2000).

24. F. Vilaplana, S. Karlsson, A. Ribes-Greus, Euro Poly J., **43**, 4371 (2007).

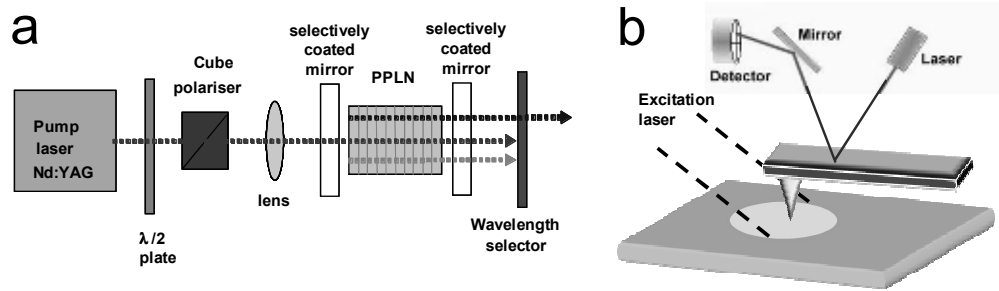

Fig. 1. Schematic drawing of a) periodically poled LiNbO3 crystal emitting tuneable

IR laser radiation, b) the experimental top-down excitation alignment.

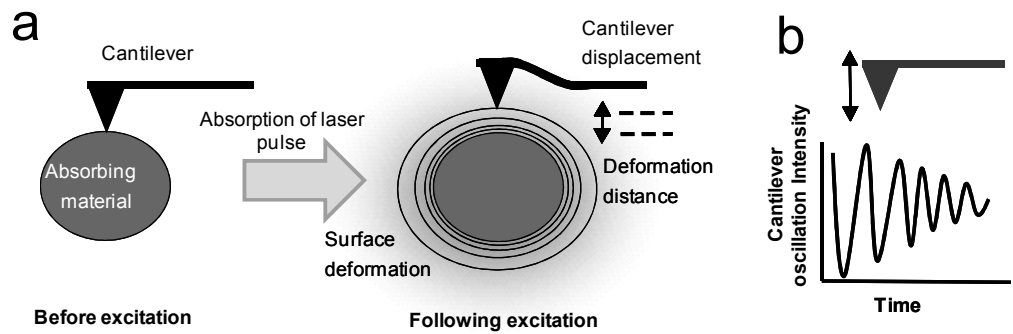

Fig 2. a) Schematic drawing of the origin of the measured signal, before on-resonance excitation, left, and the resulting surface deformation and cantilever displacement following on-resonance absorption of incident radiation, b) schematic drawing of a dampening oscillation of the cantilever following its response to the surface deformation.

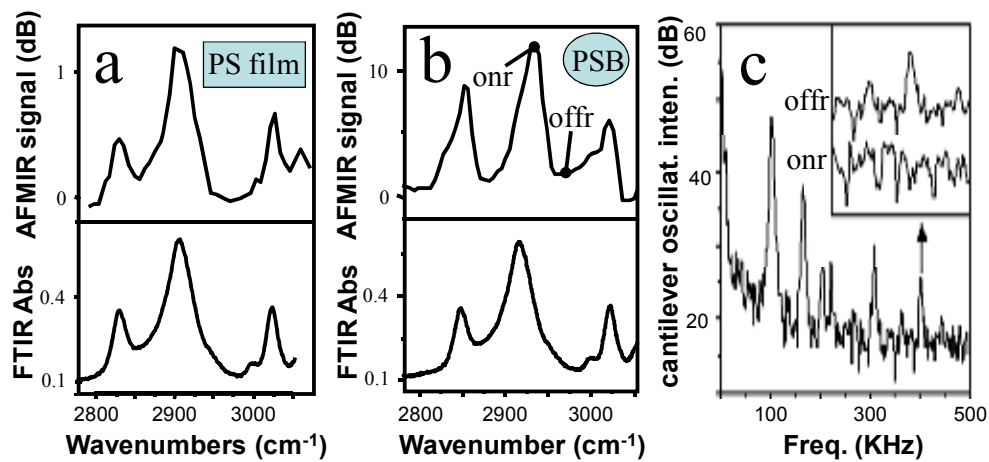

Fig 3. a) local s-AFMIR spectrum recorded with the AFM tip at a fixed position on a PS film (top), shown with a spectrum of polystyrene recorded using a commercial instrument (bottom), b) local s-AFMIR spectrum recorded with the AFM tip at a fixed position on a single PSB (top), shown with a spectrum of polystyrene recorded using a commercial instrument (bottom), c) FFT spectra of spectral plots of the cantilevers oscillations shown in a) on resonance (onr),  and off resonance (offr).

.

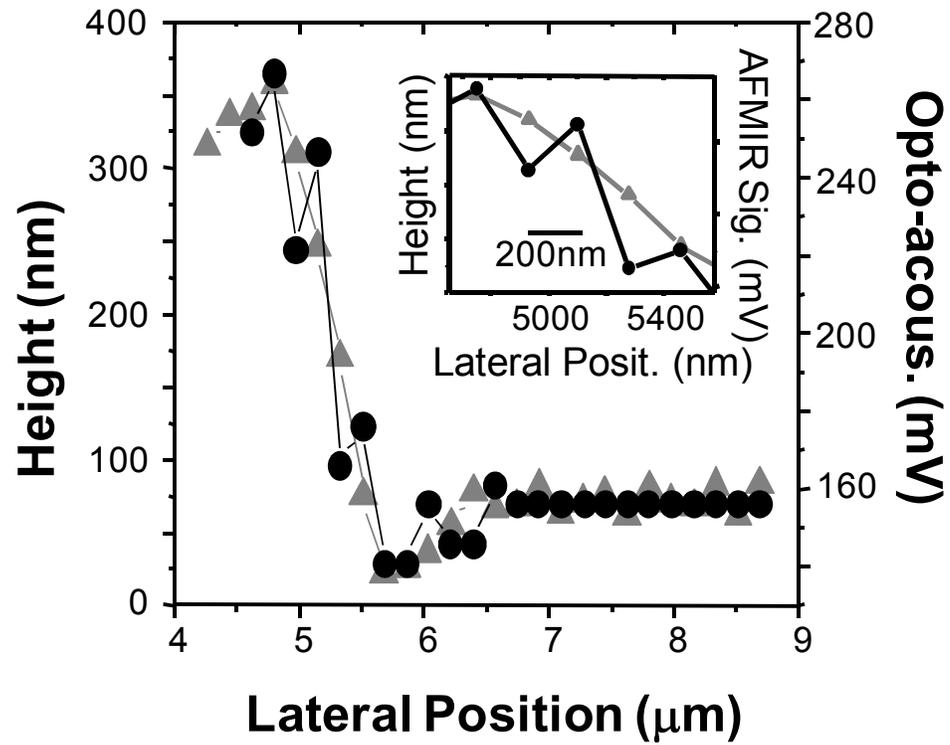

Fig. 4. a) Plot of PTIR (black curve) and AFM topographic(gray curve) data from a single PSB. Inset shows an expanded part of the plot.

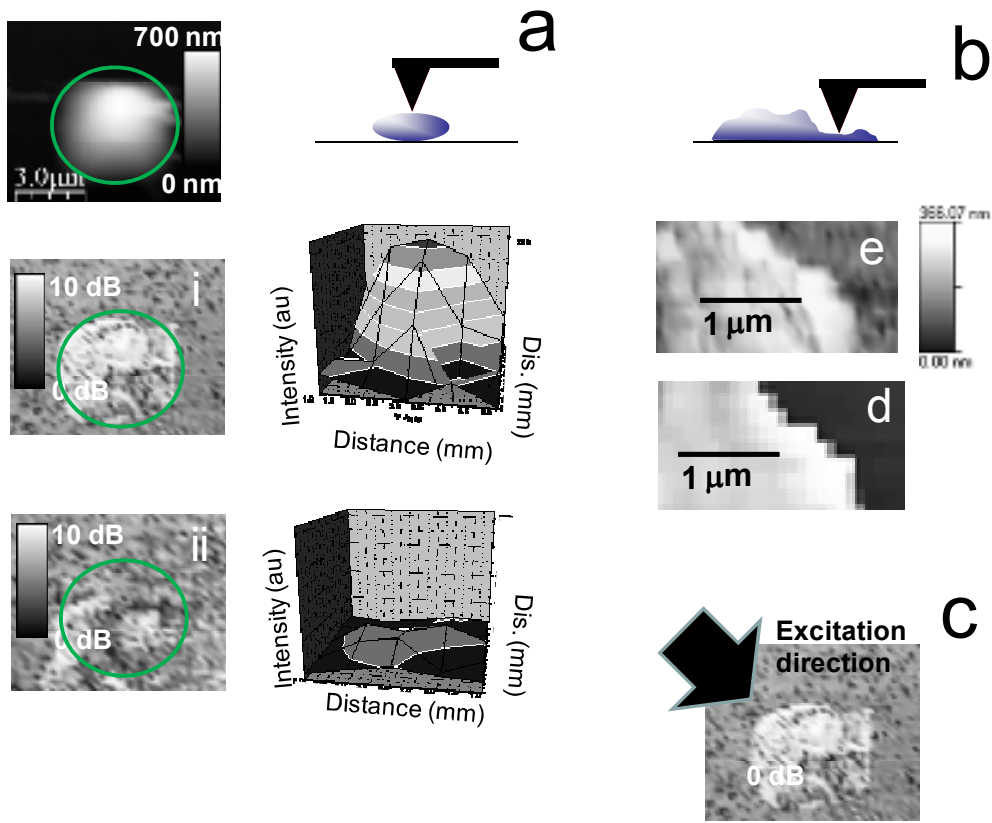

Fig. 5. a) Images recorded from a single PSB, an AFM topography image (top), localised s-AFMIR absorption images recorded at 2920 cm$^{-1}$ (on resonance) (middle)and an s-AFMIR image at 2965 cm$^{-1}$ (off resonance) (bottom), shown also are 3D plots of both s-AFMIR images shown along side their corresponding IR images. b) AFM topography (top) image and an s-AFMIR image at 2920 cm$^{-1}$ (on-resonance) for moulded PSBs. c) arrow shows the IR laser excitation direction using an image from a).

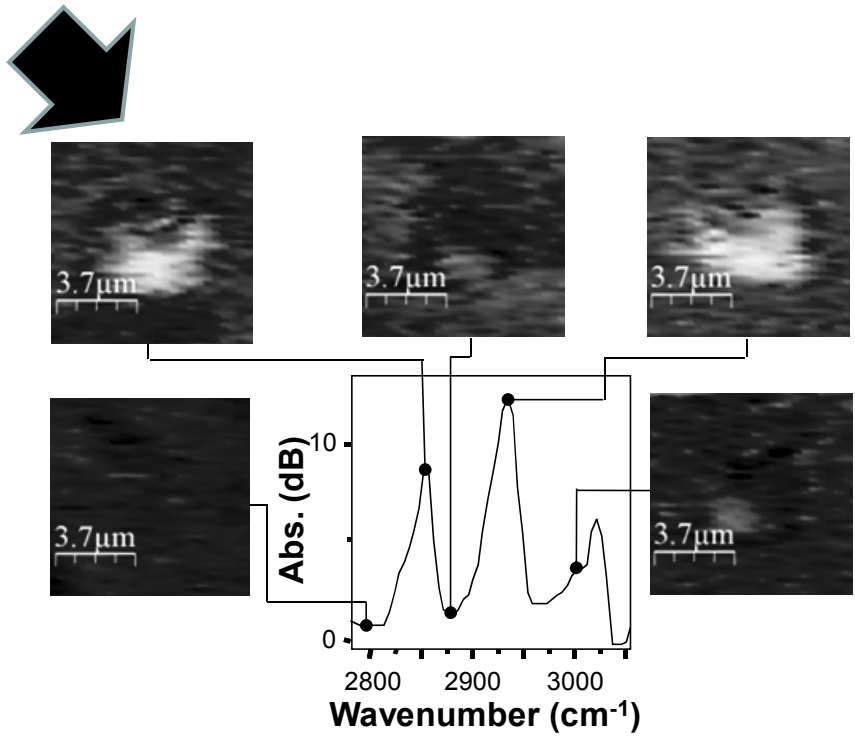

Fig. 6. s-AFMIR Images for a single polystyrene bead recorded at different excitation energies on and around the IR C-H stretch region. Arrow shows the IR laser excitation direction.